\documentstyle[prb,preprint,epsf,aps]{revtex}
\begin{document} \draft
\tightenlines
\title{ Coherent resonant tunneling in AC 
fields.}
\author{Ram\'on Aguado, Jes\'us I\~{n}arrea and Gloria Platero}
\address{Instituto de Ciencia de Materiales (CSIC) and  Departamento de 
Fisica de la Materia Condensada C-III, Universidad Autonoma,
Cantoblanco, 28049 Madrid, Spain.}
\maketitle   
\begin{abstract}
We have analyzed the tunneling transmission  probability
and electronic current density 
through resonant heterostructures in 
the presence of an external electromagnetic field.
In this work, we 
compare two different models for a double barrier :
In the first case the effect of the external field is taken into
account by
spatially dependent AC voltages 
and in the
second one the electromagnetic field is described in terms of a photon
field that irradiates homogeneously the whole sample. 
While in the first description the tunneling 
takes place mainly through photo sidebands
in the case of homogeneous illumination the 
main effective tunneling channels correspond to the coupling
between different electronic states 
due to photon absorption and emission.
The difference of tunneling
mechanisms between these configurations
is strongly reflected in the transmission and current
density which present very different features in both cases.\\
In order to analyze these effects we have obtained, 
within the Transfer Hamiltonian framework, a general expression
for the transition probability for 
coherent resonant tunneling in terms of the Green's function
of the system.
\end{abstract}
\widetext
\section{INTRODUCTION}
\par
In the last years
several works have been devoted
to the analysis of 
the effect of a time-dependent field on the transport
properties of resonant heterostructures, i.e., double
barriers, quantum wires and quantum dots \cite{1,13}. However 
there is not yet a systematic discussion of the different
situations corresponding to an AC voltage 
applied between the left and right leads
and which implies a position-dependent dephasing of
the external field on the sample , 
and the case where the whole sample 
is homogeneously illuminated ,i.e. , where
the electron-photon coupling  
 depends on position just through the momentum
matrix elements.\\
 In spite of the increasing interest in this field,
most of the theoretical work has been done
considering spatially time dependent voltages but the experimental information
includes both configurations. 
Both situations, however, are physically different 
because the effective tunneling channels 
for the flowing of current
are different in both cases.\\ 
 The coherent tunneling 
 in the presence of light applied homogeneously
on a double barrier heterostructure has been recently treated
in the scheme of the Transfer Matrix\cite{10}.\\
An alternative way to obtain the Transmission
coefficient is based in the Transfer Hamiltonian
and considers a localized basis representation, i.e.,
approximated hamiltonians whose eigenstates
are spatially localized.
The extension of the TH
 to analyze coherent resonant tunneling
(GTH) \cite{14} allows to analyze not only the sequential
tunneling which consider the
electrons tunneling through the single 
barriers, emitter and
collector, sequentially, but the coherent one
 which includes the virtual transitions
through the resonant states for electrons 
crossing coherently the
heterostructure.\\ 
In this work we have extended the Generalized Transfer
Hamiltonian formalism (GTH) \cite{14} 
to obtain 
the transition probability 
for the coherent tunneling
in the presence of a time dependent potential.\\
 In the first section of the paper we will develop
the theory to obtain the transition 
probability for a double
barrier structure in the presence 
of a spatially dependent
AC modulation . In the next section the GTH is extended to
describe the photoassisted tunneling process for a 
sample homogeneously illuminated . The different features
obtained for the coherent tunneling current
for the two different configurations are discussed in a further
section.\\
\section{TRANSITION PROBABILITY THROUGH A DOUBLE
BARRIER IN AN AC FIELD}
\par
We are going to analyze the effect of an AC field applied
just to the left and right leads with a dephasing
of $\pi$ . This 
configuration 
is schematically represented for the double barrier
in fig. 1.\\
 The Transfer Hamiltonian formalism developed by Bardeen \cite{15} ,
 allows to describe in first order time dependent perturbation theory
 the transition probability in terms of the eigenstates of
 auxiliar hamiltonians spatially localized.
 This formalism has been extended to all orders
 in perturbation theory (Generalized Transfer 
 Hamiltonian) to analyze as well tunneling
 through resonant states, i.e., to include virtual transitions
 through the localized states in the well \cite{14} (see fig.2a).\\
 The quantum mechanical Hamiltonian for an electron in the presence 
of time dependent potential can be written:
\begin{eqnarray}
H(t)=H_{L}(t)+H_{R}(t)+H_{T}(t)+H.c
\end{eqnarray} 
Where $H_{L}(t)$ and $H_{R}(t)$ are the hamiltonians for the left and
right sides respectively including the time dependent perturbation and
$H_{T}(t)$ is the coupling term which accounts for the transfer of 
electrons from the left to the
right side. 
 We are going to define now $H_{L}(t)$ and $H_{R}(t)$ as:
\begin{eqnarray}
H_{L}(t)&\equiv&\sum_{k}\epsilon_{k_{L}}(t)c_{k_{L}}^{+}c_{k_{L}}=
\sum_{k}[\epsilon_{k_{L}}+\langle
k_{L}|H_{2}(t)|k_{L}\rangle]c_{k_{L}}^{+}c_{k_{L}}\nonumber\\
H_{R}(t)&\equiv&\sum_{p}\epsilon_{p_{R}}(t)c_{p_{R}}^{+}c_{p_{R}}=
\sum_{p}[\epsilon_{p_{R}}-\langle
p_{R}|H_{2}(t)|p_{R}\rangle]c_{p_
{R}}^{+}c_{p_{R}}.
\end{eqnarray}
where $c_{k_{L}}^{+},c_{k_{L}},c_{p_{R}}^{+},c_{p_{R}}$ are creation and
destruction  operators for electron in the left- and right-hand side of
the heterostructure, respectively; $\epsilon_{k_{L}}
(\epsilon_{p_{R}}$)
are the eigenenergies for $ H^{0}_{L} $  ($ H^{0}_{R} $)
 and
 $H_{2}(t)$ is the hamiltonian which describes the AC time dependent
potential
in the leads
and  can be written as:
\begin{equation}
H_{2}(t)=V_{AC}cos\omega_{0}t
\end{equation}
In these expressions just diagonal terms are considered
in the time dependent term. The reason is that the off-diagonal
terms are zero in this case due to the orthogonality
of the eigenstates of the auxiliar hamiltonians $ H_{L}^{0}, H_{R}^{0} $.
The retarded single electron Green's function of these hamiltonians are:
\begin{eqnarray}
G_{L}^{+}(t,t')&\equiv&-\frac{i}{\hbar}\theta(t-t')\langle0|c_{k_{L}}(t)
c_{k_{L}}^{+}(t')|0\rangle=-\frac{i}{\hbar}\theta(t-t')
exp [-\frac{i}{\hbar}\int_{t'}^{t} dt_{1} \epsilon_{k_{L}}(t_{1})]\nonumber\\
&=&-\frac{i}{\hbar}\sum_{n=-\infty}^{\infty}\sum_{m=-\infty}^{\infty}
J_{n}(\frac{V_{AC}}{\hbar\omega_{0}})J_{m}(\frac{V_{AC}}{\hbar\omega_{0}})
e^{-\frac{i}{\hbar}\epsilon_{k_{L}}(t-t')}
 e^{-in\omega_{0}t} e^{im\omega_{0}t'}
\nonumber\\
G_{R}^{+}(t,t')&\equiv&-\frac{i}{\hbar}\theta(t-t')\langle0|c_{p_{R}}(t)
c_{p_{R}}^{+}(t')|0\rangle=-\frac{i}{\hbar}\theta(t-t')
exp [-\frac{i}{\hbar}\int_{t'}^{t} dt_{1} \epsilon_{p_{R}}(t_{1})]\nonumber\\
&=&-\frac{i}{\hbar}\sum_{r=-\infty}^{\infty}\sum_{s=-\infty}^{\infty}
J_{r}(\frac{V_{AC}}{\hbar\omega_{0}})J_{s}(\frac{V_{AC}}{\hbar\omega_{0}})
e^{-\frac{i}{\hbar}\epsilon_{p_{R}}(t-t')}
 e^{ir\omega_{0}t} e^{-is\omega_{0}t'}
\end{eqnarray}
where $|0\rangle$ denotes the electron vacuum in the leads,
$c_{k_{L}}^{+}(t),c_
{k_{L}}(t),c_{p_{R}}^{+}(t),c_{p_{R}}(t)$ are the creation and
destruction fermion operators in the leads evaluated in the Heisenberg
representation and we have made use of the identity:
\begin{equation}
e^{-i\frac{V_{AC}}{\hbar\omega_{0}}sin\omega_{0}t}\equiv\sum_{m=-\infty}^{\infty
}
J_{m}(\frac{V_{AC}}{\hbar\omega_{0}})e^{-im\omega_{0}t}
\end{equation}
In this equation $J_{m}$ denotes the Bessel function of integer order
m.\\
We use an interaction picture to switch on adiabatically the required
perturbations that allow us to recover the total hamiltonian H(t).
In order to apply the Generalized Transfer Hamiltonian method we choose
a hamiltonian that in its first quantized form is \cite{14}:
 \begin{eqnarray}
  H(t)=H_{L}(t)+V_{L}(t) = H_{R}(t)+V_{R}(t) 
  \end{eqnarray}
Where $ V_{L}(t) $ and $ V_{R}(t) $ can be written as:
\begin{eqnarray}
V_{L}(t)&=&V_{L} e^{\eta t}-V_{AC}\{\Theta(x-x_{2})-\Theta(x-x_{3})\}cos\omega_{0}t e^{\eta t}-2V_{AC}
\Theta(x-x_{3})cos\omega_{0}t e^{\eta t}\nonumber\\
V_{R}(t)&=&V_{R} e^{\eta t}+V_{AC}\{\Theta(x-x_{2})-\Theta(x-x_{3})\}cos\omega_{0}t e^{\eta t}+2V_{AC}
\Theta(x_{2}-x)cos\omega_{0}t e^{\eta t}
\end{eqnarray}
$V_{L}(t)$ and $V_{R}(t)$ will be considered of the same order in the
perturbative procedure and $V_{L},V_{R}$ are represented in fig 2a.
The time evolution of the wave function for the total system can be written as:\\
\begin{equation}
|\Psi(t)\rangle=f(t)\sum_{m=-\infty}^{\infty}J_{m}
(\frac{V_{AC}}{\hbar\omega_{0}})e^{-im\omega_{0}t}e^{-i\omega_{k_{L}}t}|k_{L}
\rangle+
\sum_{n=-\infty}^{\infty}J_{n}(\frac{V_{AC}}{\hbar\omega_{0}})e^{in\omega_{0}t}
\sum_{p_{R}}U_{R}(t,-\infty)e^{-i\omega_{p_{R}}t}|p_{R}\rangle
\end{equation}
This wave function must describe a particle initially on the left
side. This is satisfied by taking
$f(-\infty)=1$ and 
$U_{R}(-\infty,-\infty)=0$.
The electrons in a particular state $|k_{L}\rangle$ can in principle
evolve to any state $|p_{R}\rangle$ in the right side so that a
summation over right states is required in the expression of the wave
function.  
The time evolution operator $ U_{R}(t,-\infty)$ gives the evolution
of an electron to a right state and is determined at every order
from the time dependent Schr\"odinger equation by an expansion in a perturbation
series.We take $f(t)=f^{(0)}$ and $ U_{R}(t,-\infty)=\sum_{j=1}^{\infty}
U_{R}^{(j)}(t,-\infty) $ where j denotes the perturbation order. 
Applying the Schr\"odinger equation we obtain just to
order j the set of equations:
\begin{eqnarray}
i\hbar\sum_{n=-\infty}^{\infty}J_{n} e^{in\omega_{0}t}
\frac{\partial U_{R}^{(1)}}{\partial t}
e^{{-i\omega_{p_{R}}t}}|p_{R}\rangle  &=&
\sum_{m=-\infty}^{\infty}J_{m}
e^{-im\omega_{0}t}e^{-i\omega_{k_{L}}t}V_{L}(t)
|k_{L}\rangle.\nonumber\\
&\vdots&\nonumber\\
i\hbar\sum_{n=-\infty}^{\infty}J_{n} e^{in\omega_{0}t}
\frac{\partial U_{R}^{(j)}}{\partial t}
e^{{-i\omega_{p_{R}}t}}|p_{R}\rangle   &=&
\sum_{n_{1}=-\infty}^{\infty}J_
{n_{1}}
e^{-in_{1}\omega_{0}t}\sum_{p_{R_{1}}}e^{-i\omega_{p_{R_{1}}}t}
U_{R_{1}}^{(j-1)}
V_{R}(t)
|p_{R_{1}}\rangle (j\geq 2) .
\end{eqnarray}
These iterative equations for $U_{R}^{(j)}$ are solved by projecting them
on the state $\langle p_{R}|$ ,giving :\\
\begin{eqnarray}
U_{R}^{(1)}=&-&\frac{i}{\hbar}\{\sum_{n=-\infty}^{\infty}\sum_{m=-\infty}^{\infty}
J_{n}J_{m}
\langle p_{R}|V_{L}|k_{L}\rangle\nonumber\\
&-&\sum_{n=-\infty}^{\infty}\sum_{m=-\infty}^{\infty}
\frac{1}{2}[J_{n+1}+J_{n-1}]J_{m}
V_{AC}\langle p_{R}|\Theta(x-x_{2})-\Theta(x-x_{3})|k_{L}\rangle\nonumber\\
&-&\sum_{n=-\infty}^{\infty}\sum_{m=-\infty}^{\infty}
[J_{n+1}+J_{n-1}]J_{m}
V_{AC}\langle p_{R}|\Theta(x-x_{3})|k_{L}\rangle\}\nonumber\\
&\times&\int_{-\infty}^{t}dt_{1}
e^{i(\omega_{p_{R}}-\omega_{k_{L}}-n\omega_{0}-m\omega_{0}-i\eta)t_{1}}
\end{eqnarray}
For the physical parameters considered experimentally the amplitude of
the time dependent modulation is much smaller than the barrier height, $V_{AC}<<V_{0}$
($V_{0}$ is the height of the barriers) and then it is a good
approximation to consider:
\begin{equation}
U_{R}^{(1)}\simeq -\frac{i}{\hbar}\sum_{n=-\infty}^{\infty}
\sum_{m=-\infty}^{\infty}J_{n}J_{m}\langle p_{R}|V_{L}|k_{L}\rangle
\int_{-\infty}^{t}dt_{1}e^{i(\omega_{p_{R}}-\omega_{k_{L}}-n\omega_{0}-m\omega_{0}-i\eta)t_{1}}
\end{equation} 
Then,we proceed considering just the static part of $V_{L}(t)(V_{R}(t))$;
the time evolution operator to j-order can be written:
\begin{eqnarray}
&&U_{R}^{(j)}=(-\frac{i}{\hbar})^j
\sum_{n,m}\sum_{R_{1}}\sum_{n_{1}}....\sum_{R_{j-1}}\sum_{n_{j-1}}J_{n}J_{m}
J_{n_{1}}^{2}....J_{n_{j-1}}^{2}\langle
p_{R}|V_{R}|p_{R_{1}}\rangle....\langle p_{R_{j-2}}|V_{R}|p_{R_{j-1}}\rangle
\langle p_{R_{j-1}}|V_{L}|k_{L}\rangle\nonumber\\
&&\times\int_{-\infty}^{t}dt_{1}
e^{i(\omega_{p_{R}}-\omega_{p_{R_{1}}}-n\omega_{0}-n_{1}\omega_{0}-i\eta)t_{1}}
\int_{-\infty}^{t_{1}}dt_{2}
e^{i(\omega_{p_{R_{1}}}-\omega_{p_{R_{2}}}-n_{1}\omega_{0}-n_{2}\omega_{0}-i\eta)t_{2}}....\nonumber\\
&&\times\int_{-\infty}^{t_{j-1}}dt_{j}
e^{i(\omega_{p_{R_{j-1}}}-\omega_{k_{L}}-n_{j-1}\omega_{0}-m\omega_{0}-i\eta)
t_{j}}
\end{eqnarray}
The solution is:\\
\begin{eqnarray}
&&U_{R}^{(j)}(t,-\infty)=\sum_{n,m}\sum_{R_{1}}\sum_{n_{1}}....\sum_{R_{j-1}}
\sum_{n_{j-1}}J_{n}J_{m}\frac{e^{i(\omega_{p_{R}}-\omega_{k_{L}}-n\omega_{0}
-m\omega_{0}-ij\eta)t}}{(\omega_{p_{R}}-\omega_{k_{L}}-n\omega_{0}-m\omega_{0}
-ij\eta)}\nonumber\\
&&\times\frac{1}{(-\hbar)^{j}}\frac{J_{n_{1}}^{2}....
 J_{n_{j-1}}^{2}\langle p_{R}|V_{R}|p_{R_{1}}\rangle....
\langle
p_{R_{j-1}}|V_{L}|k_{L}\rangle}{(\omega_{p_{R_{1}}}-\omega_{k_{L}}-n_{
1}\omega_{0}-m\omega_{0}-i(j-1)\eta).... (\omega_{p_{R_{j-1}}}-\omega_{k_{L}}-n
_{j-1}\omega_{0}-m\omega_{0}-i\eta)}
\end{eqnarray}
In this procedure we have assumed that the spectral density of the electrons in
the leads is:
\begin{eqnarray}
A_{k}(\epsilon)\equiv 2\pi \sum_{n=-\infty}^{\infty}J_{n}^{2}
\delta(\epsilon-\epsilon_{k}+n\hbar\omega_{0})
\end{eqnarray}
This assumption will be correct only in the nonadiabatic regime where the external 
frequency is much larger than the inverse
resonant-tunneling time \cite{6,18}.This regime does not consider the ac 
components of the spectral densities and Green's functions (see appendix).\\ 
This allows us to write the Fourier transform of the retarded
single electron Green's function of $H_{R}$ as:  
\begin{eqnarray}
G_{R}^{+}(\epsilon)=\sum_{n=-\infty}^{\infty}\sum_{p_{R_{j}}} J_{n}^{2}
\frac{ |p_{R_{j}}\rangle
\langle p_{R_{j}}| }
{\epsilon-\epsilon_{p_{R_{j}}}+n\hbar\omega_{0}+i\eta}
\end{eqnarray}
It can be found from:
\begin{equation}
A_{k}(\epsilon)=-2Im G_{k}^{+}(\epsilon)
\end{equation} 
 And then, $U_{R}^{(j)}$ becomes in terms of the retarded Green's function:
\begin{eqnarray}
&&U_{R}^{(j)}(t,-\infty)=\sum_{n=-\infty}^{\infty}
\sum_{m=-\infty}^{\infty}J_{n}J_{m}
\frac{e^{\frac{i}{\hbar}(\epsilon_{p_{R}}-\epsilon_{k_{L}}-n\hbar\omega_{0}
-m\hbar\omega_{0}-ij\eta)t}}{(\epsilon_{p_{R}}-\epsilon_{k_{L}}
-n\hbar\omega_{0}-m\hbar\omega_{0}
-ij\eta)}\nonumber\\
&&\langle p_{R}|V_{R}G_{R}^{+}(\epsilon_{k_{L}}+m\hbar\omega_{0}+i(j-1)\eta)V_{R}
....V_{R}G_{R}^{+}(\epsilon_{k_{L}}+m\hbar\omega_{0})V_{L}|k_{L}\rangle
\end{eqnarray}
 The transition probability from left to right per unit time 
can be expressed as:
\begin{eqnarray}
&P_{RL}&=\lim_{\eta\rightarrow
0}2Re[U_{R}^{*}(t,-\infty)\frac{dU_{R}(t,-\infty)}{dt}]
\nonumber\\
\end{eqnarray}
Where $ U_{R}(t,-\infty)=\sum_{j=1}^{\infty} U_{R}^{(j)}(t,-\infty) $ 
includes the sum over all orders in perturbation theory.
In the case of a continuous spectrum there are no divergences
in the analytic expression of the Green's function so that the limit
$\eta\rightarrow 0$ of the matrix elements in $U_{R}$ can be taken
independently
from that of the fractions ,and then, we can express the time evolution
in terms of the total Green's function of
the system:
\begin{equation}
G^{+}(\epsilon)=G^{+}_{R}(\epsilon)+G^{+}_{R}(\epsilon)V_{R}G^{+}_{R}(\epsilon)
+...
\end{equation}
In the case of a discrete spectrum some generalization is
needed because the adiabatic limit $\eta \rightarrow 0$ would imply to have poles
in the Green's functions which coincide with these of the fractions 
that give the delta functions \cite {14}.\\
The expression for the time dependent transition probability  is:
\begin{eqnarray}
P_{RL}&=&\frac{i}{\hbar}\sum_{nmpq}J_{n}J_{m}J_{p}J_{q}e^{i(q\omega_{0}+p
\omega_{0}-n\omega_{0}-m\omega_{0})t}\nonumber\\
&\langle& p_{R}|V_{L}+V_{R}G^{+}(\epsilon_{k_{L}}+m\hbar\omega_{0})V_{L}|
k_{L}\rangle
\langle
k_{L}|V_{L}G^{+}(\epsilon_{k_{L}}+q\hbar\omega_{0})V_{R}+V_{L}|
p_{R}\rangle^{*}\nonumber\\
&[&PP \frac{1}{\epsilon_{p_{R}}-\epsilon_{k_{L}}-p\hbar\omega_{0}-q\hbar
\omega_{0}}-i\pi\delta(\epsilon_{p_{R}}-\epsilon_{k_{L}}-p\hbar\omega_{0}-q\hbar
\omega_{0})-\nonumber\\
&&PP \frac{1}{\epsilon_{p_{R}}-\epsilon_{k_{L}}-n\hbar\omega_{0}-m\hbar
\omega_{0}}-i\pi\delta(\epsilon_{p_{R}}-\epsilon_{k_{L}}-n\hbar\omega_{0}-m\hbar
\omega_{0})]
\end{eqnarray}
Where PP denotes the principal part and $G^{+}$ is the total Green's
function of the system.
\par
In order to obtain the stationary transition probability
we consider n=p and m=q:\\ 
\begin{eqnarray}
&P_{RL}&=\frac{2\pi}{\hbar}\sum_{n,m =-\infty}^{\infty}
J_{n}^{2}(\frac{V_{AC}}{\hbar\omega_{0}})
J_{m}^{2}(\frac{V_{AC}}{\hbar\omega_{0}})
\delta(\epsilon_{p_{R}}
-\epsilon_{k_{L}}-n\hbar\omega_{0}-m\hbar\omega_{0})\nonumber\\
&|&\langle
p_{R}|V_{L}+V_{R}G^{+}(\epsilon_{k_{L}}+m\hbar\omega_{0})V_{L}|{k_{
L}}
\rangle|^{2}
\end{eqnarray}
This formula for the transition probability is a natural 
extension
of the 
Fermi's Golden Rule formulas \cite{5,6,7} 
to analyze the coherent resonant tunneling in
terms of the total Green's function of the system.
The term which does not contain the Green's
function corresponds to first order perturbation theory,
and is the only one  appearing 
in the Transfer Hamiltonian method \cite{15},
however the term containing the total Green's function is the one which
includes  processes which involve real intermediate states 
and therefore describes correctly the coherent resonant tunneling.\\
From (21) one can obtain straightforward the Transmission
probability and the coherent tunneling current density\cite{5}.\\
In many physical cases $G^{+}$ can be calculated but, 
in general,
the main
physics of the problem can be recovered 
by using approximations to $G^{+}$.
This is the case of resonant tunneling in which the main physics is
connected with well states so that a good approximation is to substitute
$G^{+}$ by the Green's function $G_{C}^{+}$ 
of an isolated quantum well (see
fig 2b) with a selfenergy which accounts for the coupling of the well
with the continuum of states in the right and left leads.
By means of a Dyson equation the selfenergy can be expressed as:  
\begin{equation}
\Sigma(\epsilon_{k_{L}}+m\hbar\omega_{0})=V_{C}+V_{C}G_{C}^{+}(\epsilon_{k_{L}
}+m\hbar\omega_{0})V_{C}
\end{equation}
The imaginary part of the selfenergy is related with the elastic
coupling to the leads as:
\begin{equation}
\Gamma(\epsilon_{k_{L}}+m\hbar\omega_{0})=
-2Im\Sigma(\epsilon_{k_{L}}+m\hbar\omega_{0})
\end{equation}
And the energy shift of the resonances:
\begin{equation}
Re\Sigma(\epsilon_{k_{L}}+m\hbar\omega_{0})=PP\int_{-\infty}^{\infty}
\frac{d\epsilon_{q_{L}}}{2\pi}\frac{\Gamma(\epsilon_{q_{L}})}
{\epsilon_{k_{L}}-\epsilon_{q_{L}}+m\hbar\omega_{0}}
\end{equation}
is the Hilbert Transform of the elastic coupling.\\ 
With this procedure the energy dependence of
the broadening of the resonant levels for each bias voltage
applied 
across the heterostructure (this dependence is included in the
potentials $V_{C}$) is considered in a straightforward way.\\
Also its dependence on the external field is included through the
dependence of $\Sigma$ on the energy of the m photo sideband 
(ie., the broadening of each photo sideband is different ) 
and through the spectral densities (see formula 14)
that give the correct weight to each photo sideband.
One should remark that usually
the models based in the Transfer Hamiltonian define the elastic coupling 
to the leads as :
\begin{eqnarray}
\Gamma(\epsilon)&=&\Gamma_{L}(\epsilon)+\Gamma_{R}(\epsilon)\nonumber\\
&=& \sum_{k_{L}}T_{k_{L}}^{2}A_{k_{L}}(\epsilon)
+\sum_{p_{R}}T_{p_{R}}^{2}A_{p_{R}}(\epsilon) 
\end{eqnarray}
Here $ T_{k_{L}},T_{p_{R}}$ are the matrix elements 
given by Bardeen \cite{15}
and $A_{k_{L}},A_{p_{R}}$ are the spectral densities of the left and
right leads respectively .
This expression implies that the transmission probability is the  sum
of the transmission probabilities of each barrier separately and
therefore that the tunnel is sequential \cite{5,6,12,13}.
We consider coherent tunnel
and this is the reason to define the broadening
as in Eq(22-23). Therefore, by means of our model 
we do not have to treat the
coupling with  the leads as a constant as the usual Transfer
Hamiltonian methods for sequential tunnel do.
\par

\section{ LASER IRRADIATING
THE WHOLE HETEROSTRUCTURE}
\par
We consider now the case where the light illuminates homogeneously
the whole heterostructure.
In this case, there is no a position-dependent
phase shift of the external field on the heterostructure and the
spatial dependence, as we will see below , appears
through the matrix elements of the 
electronic momentum operator.\\
The electromagnetic field is represented by a plane electromagnetic wave
of wave vector $\vec{k}$, parallel to the $x$ direction and polarized in
the tunnel direction (see fig.3)
$\vec{E}=(0,0,F)$.
The  hamiltonian for this configuration has been solved 
within the framework of the Transfer
Matrix formalism and time dependent 
perturbation theory.\cite {10}\\
Our aim now is to obtain the transition 
probability and the tunneling
current
within the scheme of the Transfer 
Hamiltonian, in the same way as it was
done in the previous section for the AC case. 
The interest of doing
that in this case is again to work 
in terms of localized states
as basis and therefore it would allow 
to consider, as it
was discussed previously, 
different kind of excitations in the
different spatial regions, as well as systems with
localized eigenstates.\\
Applying the same procedure as 
in the case of the oscillating voltages we can
define the hamiltonians $H_{L}(t)$ and $H_{R}(t)$
including the time dependent
part of the hamiltonian that includes only diagonal coupling between
electronic states.In the Coulomb Gauge :
\begin{eqnarray}
H_{L}(t)&=&\sum_{k_{L}}[\epsilon_{k_{L}}+\frac{e}{
m^{*}}
\langle k_{L}|P_{z}|k_{L}\rangle
(\frac{\hbar}{2\epsilon
V\omega_{0}})^{\frac{1}{2}}
(ae^{-i\omega_{0}t}+a^{+}e^{i\omega_{0}t})]c_{k_{L}}^{+}c_{k_{L}}\nonumber\\
H_{R}(t)&=&\sum_{p_{R}}[\epsilon_{p_{R}}+
\frac{e}{m^{*}}\langle p_{R}|P_{z}|p_{R}\rangle
(\frac{\hbar}{2\epsilon
V\omega_{0}})^{\frac{1}{2}}
(ae^{-i\omega_{0}t}+a^{+}e^{i\omega_{0}t})]c_{p_{R}}^{+}c_{p_{R}}
\end{eqnarray}
Here the vector potential operator is:
\begin{equation}
\vec{A}=(\frac{\hbar}{2\epsilon
V\omega_{0}})^{\frac{1}{2}}(ae^{-i\omega_{0}t}+
a^{+}e^{i\omega_{0}t})\vec{\epsilon_{z}}
\end{equation}
$a^{+}$,$a$ are the creation and destruction operators for photons and
the
wave vector of the electromagnetic field has been neglected.
We will see below that the comparison with the previous
configuration indicates that
there are qualitative features
in the transmission coefficient and
current density which are very different for both
cases. Therefore the configuration
of  the time dependent potential matters for analyzing
the photoassisted transport properties.\\
The main reasons for those differences are,
firstable, that in this case,
 the electron-photon coupling term contains
the $\vec{P}$ operator matrix elements , and secondly,
that there are off-diagonal terms in
the Hamiltonian, i.e.,
terms which describe how the electromagnetic
field couples the different electronic states.
These terms are those which produce a difference in
the transition probability with respect to the
case without a time dependent field applied to
the sample.
In the previous case, however, the transition probability
difference comes from the photo sidebands which appear
in the regions affected by the time dependent field and
which behave as additional tunneling channels.\\
The left and right hamiltonians are exactly soluble, by means
of a canonical transformation \cite{10,16}  
which decouples electrons and photons.
We will not repit here this procedure 
but we give the  expression for
the retarded Green's functions associated 
with the hamiltonians $H_{L}(t)$ and
$H_{R}(t)$:
\begin{eqnarray}
G_{L}^{+}(t,t')
&=&-\frac{i}{\hbar}\sum_{n=-\infty}^{\infty}\sum_{m=-\infty}^{\infty}
J_{n}(\beta_{k_{L}})J_{m}(\beta_{k_{L}})
e^{-\frac{i}{\hbar}\epsilon_{k_{L}}(t-t')}
 e^{-in\omega_{0}t} e^{im\omega_{0}t'}
\nonumber\\
G_{R}^{+}(t,t')
&=&-\frac{i}{\hbar}\sum_{r=-\infty}^{\infty}\sum_{s=-\infty}^{\infty}
J_{r}(\beta_{p_{R}})J_{s}(\beta_{p_{R}})
e^{-\frac{i}{\hbar}\epsilon_{p_{R}}(t-t')}
 e^{ir\omega_{0}t} e^{-is\omega_{0}t'}
\end{eqnarray}
where:
\begin{equation}
\beta_{k}=\frac{eF\langle k|P_{z}| k \rangle}{m^{*}\hbar\omega^{2}_{0}}
\end{equation}
Here, the argument of the Bessel function 
depends on the matrix
element of the momentum operator
and its value is very small. It is also
a significative difference with respect to 
the previous case where the argument of the Bessel functions
can be arbitrarily big (depending of the
amplitude of the external field) 
and therefore the contribution of the terms which correspond
to
Bessel functions of higher order than zero and one, have to
be considered.\\
In the present case we are discussing, as the argument of 
the Bessel functions is close to zero for typical intensities and
frequencies of the field , without lose of generality it
is usually a very good approximation 
to neglect terms of order higher
than zero or one at most.\\
In order to obtain the transition probability
for coherent tunneling through the double barrier structure,
we proceed evaluating,
as in the previous case , and in the  framework 
of the Generalized Transfer Hamiltonian\cite{14},
 the time evolution operator with:
\begin{eqnarray}
V_{L}(t)=V_{L}e^{\eta t}\nonumber\\
V_{R}(t)=V_{R}e^{\eta t} 
\end{eqnarray}
Its expression,
obtained up to infinite order in time dependent perturbation
theory for the barriers and exact for the diagonal
terms in the electron-photon coupling, is the same as 
eq.(17) but
now the argument of the Bessel functions depends
on the electronic momentum
matrix elements as was already discussed.\\
At this point we include the non-diagonal coupling terms in order to
obtain the total time evolution operator. Its expression in interaction
representation is :
\begin{eqnarray}
U(t,-\infty)=1&-&\frac{i}{\hbar}\int_{-\infty}^{t}ds
\hat{H_{2}}(s)\nonumber\\
&+&(-\frac{i}{\hbar})^{2}\int_{-\infty}^{t}\int_{-\infty}^{s}d\tau
\hat{H_{2}}(s)U(s,\tau)\hat{H_{2}}(\tau)
\end{eqnarray}
Where $H_{2}$ is :
\begin{eqnarray}
H_{2}(t)&=&\sum_{k_{i}}\sum_{k^{'}_{i}}\frac{eF}{m^{*}\omega_{0}}
\langle k^{'}_{i}|P_{z}|k_{i}\rangle
c_{k^{'}_{i}}^{+}c_{k_{i}}
cos(\omega_{0}t)\nonumber\\
i &=&L,R
\end{eqnarray}
The caret on the $H_{2}$ operators indicates that  they should be evaluated
in interaction representation (here the "non interacting" hamiltonian is
just this one we have solved exactly by means of the GTH) .\\
We consider only first order and we keep only the $J_{0}$ Bessel
function terms because they are the only ones that give non negligible
contributions due to the smallness of their arguments.\\ 
The new term appearing in the time evolution operator has the
expression:
\begin{eqnarray}
&&U_{2}(t,-\infty)=\frac{1}{\hbar}\frac{J_{0}(\beta_{p_{R}})J_{0}(\beta_{k_{L}})}{2}
\frac{eF}{m^{*}\omega_{0}}
\frac{e^{i(\omega_{p_{R}}-\omega_{k_{L}}+\omega_{0}-i\eta)t}}{(\omega_{p_{R}}-
\omega_{k_{L}}+\omega_{0}-i\eta)}\nonumber\\
&&\{\langle
p_{R}|V_{R}G^{-}_{L}(\epsilon_{p_{R}}) P_{z}|
k_{L}\rangle+ \langle
p_{R}|P_{z}G^{+}_{R}(\epsilon_{k_{L}})V_{L}|
k_{L}\rangle+\nonumber\\
&&\langle
p_{R}|V_{R}G^{-}(\epsilon_{p_{R}})V_{L}G^{-}_{L}(\epsilon_{p_{R}}) P_{z}|
k_{L}\rangle+\langle p_{R}|P_{z}G^{+}_{R}(\epsilon_{k_{L}})V_{R}G^{+}
(\epsilon_{k_{L}}) V_{L}|k_{L}\rangle\nonumber\\
&&\langle p_{R}|V_{R}G^{-}(\epsilon_{p_{R}})P_{z}G^{+}(\epsilon_{k_{L}}) V_{L}|
k_{L}\rangle\}+(\omega_{0}\rightarrow -\omega_{0})
\end{eqnarray}
Where $G_{R}^{\pm}$ and $G_{L}^{\pm}$ are the Green's functions of $H_{R}$ and
$H_{L}$ respectively, $G^{\pm}$ includes the tunneling terms and we have used
the property that the total Green's function of the system can be
expanded in terms of the two  basis left and right :
\begin{equation}
G=G_{L}+G_{L}V_{L}G_{L}+....=G_{R}+G_{R}V_{R}G_{R}+...
\end{equation}
The Green's functions appearing in this expression are:
\begin{eqnarray}
G_{L}^{\pm}(\epsilon)&=&\sum_{k'_{L}} J_{0}^{2}(\beta_{k'_{L}})
\frac{ |k'_{L}\rangle
\langle k'_{L}| }
{\epsilon-\epsilon_{k'_{L}}\pm i\eta}\nonumber\\
G_{R}^{\pm}(\epsilon)&=&\sum_{p'_{R}} J_{0}^{2}(\beta_{p'_{R}})
\frac{ |p'_{R}\rangle
\langle p'_{R}| }
{\epsilon-\epsilon_{p'_{R}}\pm i\eta}\nonumber\\
G^{\pm}(\epsilon)&=&\sum_{q_{C}} J_{0}^{2}(\beta_{q_{C}})
\frac{ |q_{C}\rangle
\langle q_{C}| }
{\epsilon-\epsilon_{q_{C}}-i\frac{\Gamma(\epsilon)}{2}\pm i\eta}
\end{eqnarray}
Following the same steps as in section II the transition probability can
be written as a Fermi's golden rule:
\begin{equation}
P_{RL}=\frac{2\pi}{\hbar}\{|A_{RL}|^{2}\delta(\epsilon_{p_{R}}-\epsilon_{k_{L}})
+|B_{RL}|^{2}\delta(\epsilon_{p_{R}}-\epsilon_{k_{L}}+\hbar\omega_{0})
+(\omega_{0} \rightarrow -\omega_{0})\}
\end{equation}
where $A_{RL}$ and $B_{RL}$ contain the matrix elements:
\begin{eqnarray}
A_{RL}&=&J_{0}(\beta_{p_{R}})J_{0}(\beta_{k_{L}})
\langle p_{R}|V_{L} + V_{R}G^{+}(\epsilon_{k_{L}})V_{L}|k_{L}\rangle\nonumber\\
B_{RL}&=&\frac{J_{0}(\beta_{p_{R}})J_{0}(\beta_{k_{L}})
}{2}
\frac{eF}{m^{*}\omega_{0}}
\{\langle p_{R}|V_{R}G^{-}_{L}(\epsilon_{p_{R}}) P_{z}|
k_{L}\rangle+ 
\langle p_{R}|P_{z}G^{+}_{R}(\epsilon_{k_{L}})V_{L}|
k_{L}\rangle+\nonumber\\
&&\langle p_{R}|V_{R}G^{-}(\epsilon_{p_{R}})V_{L}G^{-}_{L}(\epsilon_{p_{R}})
P_{z}|
k_{L}\rangle+\langle p_{R}|P_{z}G^{+}_{R}(\epsilon_{k_{L}})V_{R}G^{+}
(\epsilon_{k_{L}}) V_{L}|k_{L}\rangle\nonumber\\
&&\langle p_{R}|V_{R}G^{-}(\epsilon_{p_{R}})P_{z}G^{+}(\epsilon_{k_{L}})
V_{L}|
k_{L}\rangle\}
\end{eqnarray}
\par
In the previous expression (37) the different terms contain the 
possible transitions from the left to the right states
assisted by the light. Both resonant transitions
through the well states and non resonant ones, for coherent
tunneling are included.This kind of matrix elements have been obtained
in ref 7 in the context of sequential photoassisted tunneling.\\
\section{ RESULTS FOR THE TWO DIFFERENT CONFIGURATIONS}
\par
In this section we have studied the two different 
configurations discussed in the previous section
i.e., the transmission coefficient for electrons through 
a double barrier in the presence of 
a time dependent
potential applied between the emitter and collector whose amplitude
is position-dependent and an homogeneous electromagnetic
field irradiating the whole sample .
In fig.(4) the transmission coefficient for the first case 
has been drawn. 
The heterostructure consists in a double barrier of
$GaAs-Al_{x}Ga_{1-x}As $ with well and barrier thicknesses
of 50 $ \AA $ and 100 $ \AA $ respectively.
We have considered that the energy of
the time-dependent field is 
13.6 meV corresponding to a frequency of 3.3 THz. The important magnitude in this case is
the ratio between the amplitude and the energy of the field which
enters in the argument of the Bessel functions (see formula 5).
As we have already discussed depending of the value of this number
higher order Bessel functions are non negligible
and its value determines the number of them (i.e., the 
number of photo sidebands) which participate in the transmission
probability and current density. In fig.(4) we consider that
this ratio is one half and we have 
included in the calculation the Bessel functions up 
to fourth order. This fact is reflected in the four satellites
which appear at both sides of the main central peak (which 
corresponds to the Bessel function of zero-order).\\
One should also remark that the broadening of the resonant states
 in the well
is not a constant but depends 
on the photo sideband index m (see formula
23).
In this case we observe that the contribution 
of the photo sideband of index  $ m > $0
to the transmission coefficient is
smaller than the one coming from
the main peak (m=0) but of the same order of
magnitude even for the higher order side-bands due
to the finite value of the
argument of the Bessel function  
$\frac{V_{AC}}{\hbar\omega_{0}} $.
Another interesting effect which was already discussed
by M. Wagner\cite{17} is the quenching of the transmission probability 
in the presence of an AC field.
This effect can be understood looking to formula (21) 
where the transition probability at the
energy of the resonant state is weighted by :
 $ \sum_{n=-\infty}^{\infty} 
 J_{0}^{2}  J_{n}^{2} $.
The quenching of the transmission takes place
if the argument of the
Bessel function $\beta$ = $\frac{V_{AC}}{\hbar\omega_{0}}$
is such that $J_{0}(\frac{V_{AC}}{\hbar\omega_{0}})=0 $.
In fig (5) this effect is shown. We observed it 
for different values of $ \beta $ : as it 
approaches to the complete quenching limit  the resonant 
transmission main peak is strongly reduced, therefore
tuning the intensity and frequency of the time dependent modulation 
allows to reduce the transmission independently
of the transparency of the barriers \cite{17}.\\
We have also plotted the current density for an
AC potential through a double barrier (fig.6) for different amplitudes
: 
As main features we observe that the threshold  tunneling current
moves to lower bias and also that for higher
bias the current density is smaller in the presence of the AC field,
and this effect increases as the ratio $\frac{V_{AC}}{\hbar\omega_{0}}$
increases too. Also a step-like behaviour is
observed in the
current as a function of the external bias.
We can explained these features in terms of the photo sidebands:
the threshold, for this case is close to zero bias. That is due to the
fact that 
there are photo sidebands associated to electronic states close to
the Fermi energy which contribute 
to the resonant tunneling even when
the resonant state ( $ E_{r} $ )
 is higher in energy than the emitter Fermi energy $ E_{F} $. 
If $ E_{r} $ is higher than  $ E_{F} $ in several photon
energies the photo sidebands which allow
the flow of current have a low spectral
weight and the contribution to the current is small. As far as the
resonant state energy closes into the Fermi energy, increasing the
bias, the lower indexes photo sidebands,
i.e., those which are more intense in the 
spectral function can be aligned with
the resonant state, and therefore, their contribution
to the current increases. Once the resonant state crosses the Fermi
energy the current density increases but remains 
smaller than the current with no AC applied.
That is due again to the fact that the spectral function has 
finite weight in all 
the photo sidebands and not only in the
main one which is weighted by the zero
order Bessel function and whose value is smaller than 
one. In this case , only 
a small number of them , for a fixed bias
tunnels resonantly and the effect of the field is
to reduce the current.\\
As the ratio between the field intensity and
the frequency increases it does the coupling
of the electron motion with the external field and the
current density differs more from the
case without this interaction and presents abrupt steps
coming from the satellite bands .\\
\par
The homogeneous illumination of the sample has been analyzed for the first
time
in ref 10. In order to compare this situation with the one
discussed previously, we have plotted the transmission coefficient
for a GaAs-AlGaAs double barrier for both configurations (fig.7).To make
the two situations comparable we estimate the 
ratio $\frac{V_{AC}}{\hbar\omega_{0}}$ supposing the total potential
drop due to the light to be $Fd$ where $d$ is the length of the
heterostructure, for the values of intensity and energy of the
electromagnetic field studied we have 
$\frac{eFd}{\hbar\omega_{0}}=0.77$.
The main difference
between the two configurations
 is that for these heterostructures
the momentum matrix elements are very small, 
therefore, in the case of homogeneous
light irradiating the sample and for typical 
values of the intensities and frequencies of
the electromagnetic field, the argument of the Bessel functions
of higher order than zero are negligible and therefore 
the intensity of the corresponding photo sideband can
be neglected.
That is the reason why to consider just the main peak
(the m=0 term)  makes sense. \\
However we observe in fig.7b
 the presence of two satellites in the transmission
coefficient.
These side-peaks have another 
origin than the photo sidebands \cite{10}:
They come from the mixing of electronic states due to
the homogeneous light and which show up in the hamiltonian
as the off-diagonal matrix elements of the electronic
momentum coupled by the light.
In this case just processes involving
absorption and emission of one photon are considered
 because they are the main ones \cite{10}.\\
Therefore , the tunneling channels for the two configurations are
different: in the case of an AC field 
the off-
diagonal terms cancel if the amplitude of the AC potential
is considered constant within each
region (left, center and right). In this case, the
main tunneling channels (the only ones within this approximation) 
 are the  sidebands: those in the emitter 
align in energy with the sidebands
in the well producing additional contributions
to the transmission probability and the resonant
current. Their contribution
can be important even for high order photo sidebands if
the ratio $\frac{V_{AC}}{\hbar\omega_{0}}$ is of the order
or higher than one.\\
In the case of an homogeneous electromagnetic field , 
the situation is different.
The off-diagonal electron-photon coupling terms in the hamiltonian
are  those which modify the
current density. Those channels, involving different
electronic states (fig.7 ) contribute in principle also
with all
 their photo sidebands , however, as the argument of the Bessel 
functions ( i.e., the intensity of the m  photo sideband ) 
is controlled by the momentum matrix elements 
and remains very small, 
just the zero index photo sideband (the main one)
is non-negligible and gives a contribution to the
current. Therefore the three peaks in the transmission 
coefficient come from the main bands (index
zero) corresponding to three electronic states which differ
in one photon energy and which are mixed by the
field.\\
In fig. 8.a the current density through the double barrier
 in the presence of homogeneous light
is represented. For the field intensity and frequency considered: 
$F=4.10^{5} \frac{V}{m} ,\hbar w_{0}=13.6 meV $ (3.3 THz) 
the effect of the external field is very small  due to
the electron-photon coupling term 
involving the momentum matrix elements,
and the effect 
of the light cannot be observed 
in the characteristic curve but in the
current density difference with respect to the case of no light
present in the sample (fig. 8.b) \cite{10}.\\
This result is in very good agreement with the experimental results of
ref 3 where a double barrier is irradiated with a laser in the far
infrared regime \cite{3,10} .
\par
\section{CONCLUSIONS}
\par
In this work we have extended the Generalized Transfer Hamiltonian
which allows to describe the coherent tunneling through a resonant
heterostructure\cite{14}, to include the effect of a time dependent 
field. Two different configurations have been addressed: an homogeneous 
electromagnetic field applied to the heterostructure and a 
position-dependent AC field dropping through the heterostructure.\\
We have obtained the transition probability and the current density 
through a double barrier structure for both configurations, and
the results have been discussed in terms of the 
main available tunneling
channels for each situation.\\
In the first case the photoassisted tunneling is mainly due to the 
coupling between different electronic states due to photon
absorption and emission processes. 
In this case, the satellite bands with
index higher than zero have negligible 
contribution to the transmission probability.\\
In the second one the 
sidebands at the emitter region carry the electronic charge by aligning
with those corresponding to well and collector electronic states.
The intensity
and frequency of the AC field determines the number of effective
channels (sidebands) participating in the current.\\
We did not include the effect of the electron-electron interaction
in this formalism. This effect  is small in double barrier
heterostructures but it has to be included 
for systems where the correlation is an important contribution. 
 This is the purpose of a future work.\\
\section{APPENDIX}
\par 
In this appendix we want to derive the expression for the spectral densities
and to discuss in which limit is correct to apply the Tien and Gordon's
expression for the electronic density of states \cite{1}.
We start with the retarded Green's function which depends
explicitly on two times t and t' but not on their difference 
because of the breakdown of
time translational invariance due to the external field : 
\begin{eqnarray}
G_{k}^{+}(t,t')&\equiv&-\frac{i}{\hbar}\theta(t-t')\langle0|c_{k}(t)
c_{k}^{+}(t')|0\rangle=-\frac{i}{\hbar}\theta(t-t')
exp [-\frac{i}{\hbar}\int_{t'}^{t} dt_{1} \epsilon_{k}(t_{1})]\nonumber\\
&=&-\frac{i}{\hbar}e^{-\frac{i}{\hbar}\epsilon_{k}(t-t')}
e^{-\frac{i}{\hbar}\frac{V_{AC}}{\omega_{0}}sin\omega_{0}t}
e^{\frac{i}{\hbar}\frac{V_{AC}}{\omega_{0}}sin\omega_{0}t'}\nonumber\\
&=&-\frac{i}{\hbar}\sum_{n=-\infty}^{\infty}\sum_{m=-\infty}^{\infty}
J_{n}(\frac{V_{AC}}{\hbar\omega_{0}})J_{m}(\frac{V_{AC}}{\hbar\omega_{0}})
e^{-\frac{i}{\hbar}\epsilon_{k}(t-t')}
 e^{-in\omega_{0}t} e^{im\omega_{0}t'}
\end{eqnarray}
Let us now calculate the spectral density from the retarded Green's function
which in Wigner coordinates 
$\tau=t-t' , T=\frac{t+t'}{2}$ 
is defined as \cite{19} :
\begin{equation}
A_{k}(\omega,T)=\int_{-\infty}^{\infty}d\tau e^{i\omega\tau} 
G_{k}(T+\frac{\tau}{2},T-\frac{\tau}{2})
\end{equation} 
and:
\begin{equation}
G_{k}(T+\frac{\tau}{2},T-\frac{\tau}{2})=i[G^{+}_{k}(T+\frac{\tau}{2},T-\frac{\tau}{2})
-G^{-}_{k}(T+\frac{\tau}{2},T-\frac{\tau}{2})]
\end{equation}
Performing the Fourier transform in the relative time we obtain:
\begin{equation}
A_{k}(\omega,T)=\frac{2\pi}{\hbar}\sum_{n=-\infty}^{\infty}\sum_{m=-\infty}^{\infty}
J_{n}(\frac{V_{AC}}{\hbar\omega_{0}})J_{m}(\frac{V_{AC}}{\hbar\omega_{0}})
cos[(n-m)\omega_{0}T] \delta(\omega-\omega_{k}
+\frac{n+m}{2}\omega_{0})
\end{equation}
Only in the nonadiabatic regime when the frequency of the external field is
larger than the inverse of the tunneling time for the electrons we can define an
averaged spectral density as:
\begin{equation}
A_{k}(\omega)=\frac{1}{T_{0}}\int_{0}^{T_{0}} dT A_{k}(\omega,T)
=\frac{2\pi}{\hbar}
\sum_{n=-\infty}^{\infty}J_{n}^{2}(\frac{V_{AC}}{\hbar\omega_{0}})
\delta(\omega-\omega_{k}+n\omega_{0})
\end{equation}
where the average is taken over one period $T_{0}$ of the external field.\\
With this procedure we restrict ourselves to the dc component of the spectral
density (and then of the Green's functions) since the ac components (equation 41) are
suppressed in the average.\\

\section{ACKNOWLEDGMENTS}
\par
One of us (Ram\'on Aguado) acknowledge the Universidad Carlos III de Madrid
for financial support. We acknowledge Dr.Tobias Brandes for helpful discussions
and a critical reading of the manuscript,
 also we acknowledge Prof.Carlos Tejedor and Dr.Luis Brey for enlighting discussions.\\
 This work has been supported in part by the Commission Interministerial
de Ciencia y Tecnologia of Spain under contract MAT 94-0982
-c02-02 by
the Commission of the European Communities under contract SSC-CT 90 0201
and by the Acci\'on Integrada Hispano-Alemana HA84.\\

\newpage
\begin{figure}
\caption{Schematic drawing of a resonant-tunneling structure with 
AC voltages applied in the leads.}
\end{figure}
\begin{figure}
\caption{Sketch of the different Hamiltonians used to study tunneling
with the GTH method.(a)Left and right Hamiltonians,(b)center
Hamiltonian.}
\end{figure}
\begin{figure}
\caption{Schematic drawing of a resonant-tunneling structure in the
presence of an electromagnetic field polarized in the tunnel direction.}
\end{figure}
\begin{figure}
\caption{$Log_{10}$ of coherent  transmission coefficient as a function
of
energy with and without an AC field of parameters $\frac{V_{AC}
}{\hbar\omega_{0}}=0.5, \hbar\omega_{0}=13.6 meV$ for a GaAs/AlGaAs
double barrier 100-50-100 $\AA$.}
\end{figure}
\begin{figure}
\caption{Quenching of the  transmission
for $\frac{V_{AC}}
{\hbar\omega_{0}}=2.0,
 \frac{V_{AC}}{\hbar\omega_{0}}=2.4,
 \frac{V_{AC}}{\hbar\omega_{0}}=2.5,
 \hbar\omega_{0}=13.6 meV $ (sample fig4).}
\end{figure}
\begin{figure}
\caption{ Coherent tunneling current density as a function of voltage
for the sample of fig4. 
(a)With and without an AC field of parameters
$\frac{V_{AC}}{\hbar\omega_{0}}=0.5, \hbar\omega_{0}=13.6 meV$. 
(b)For different ratios between the intensity and the energy of
the AC field ( $\hbar\omega_{0} =13.6 meV$).}
\end{figure}

\begin{figure}
\caption{Comparison  of $Log_{10}$ of coherent  transmission coefficient
as a function of 
energy for an AC field and an electromagnetic field. 
(a)$\frac{V_{A
C}
}{\hbar\omega_{0}}=0.77.
 (b) F=4.10^{5}\frac{Volt}{m},  \hbar\omega_{0}=13.6
meV$}
\end{figure}
\begin{figure}
\caption{(a)Current Density in the presence of an electromagnetic field
for the same parameters as fig7.b
(b)Current density difference.}
\end{figure}
\
\end{document}